\documentclass[12pt,.worldsci]{article}
\begin{document}
%\begin{flushright}
%UH-511-844-96  \\
%February 1996
%\end{flushright}
\thispagestyle{empty}
\textwidth 185mm
%Draft Outline Revised February 26, 1996
\begin{center}
\vglue 3.5cm
{\large \bf SONOLUMINESCENCE IN NEUTRON STARS}\\
\vspace{6.0ex}
{\large Walter Simmons, John Learned, Sandip Pakvasa, and Xerxes Tata}\\
\vspace{1.5ex}
{\large \it Department of Physics and Astronomy, University of Hawaii,\\
Honolulu, HI 96822, U.S.A}\\
\vspace{1.5ex}
%{\large Collaboration}\\
\vspace{2.0cm}
\begin{abstract}
After a brief discussion of a possible relationship between
the electroweak phase transition in highly compressed matter
and gravitational collapse, we examine the speculative possibility
that the electroweak phase transition
might be contemporarily occurring in processes in neutron stars.
We conjecture that adiabatic compression of neutron star matter
due to focusing of the energy from a supernova bounce into a
very small volume could result in extreme densities, and Fermi
levels or temperature above $\cal{O}$ (100 GeV). We propose a 
qualitative scenario for sonoluminescence in neutron stars and discuss 
possible observable consequences.
\end{abstract}
\end{center}

\section{Introduction}

In the past few decades, a description of all fundamental 
processes, except gravity, has emerged and is referred to as the 
Standard Model of particle physics[1].  According to be the Model, baryon
number and lepton number non-conservation occur, but are extremely strongly
suppressed except in processes which take place at very high temperature
or under conditions of very high density.  The suppression
is so strong that for practical purposes baryon and lepton number may be
regarded as conserved.
The theory engenders a phase
transition, the electro-weak phase transition, which takes place when the
characteristic temperature exceeds the Higgs vacuum expectation value 

\[
v\sim [\sqrt{2}G_F]^{-\frac 12}\sim 250 \ \mbox{GeV} 
\]
where $G_F$ is the Fermi constant. For temperatures in this range and above, 
baryon and lepton number non-conservation become important.
	At still higher temperature, above the sphaleron energy, which is,
\[
E = \frac{2M_W}{\alpha_W} B (\lambda/g^2) \sim 8-15 \ TeV,
\]
(where $M_W$ is the $W$ boson mass, and the function B depends upon 
the Higgs mass via the Higgs self coupling $\lambda$ and $g$ the gauge 
coupling constant, and
$\alpha_W = g^2/4 \pi$),
the Model predicts that baryon and lepton number violation is
completely unsuppressed\cite{MAT}.

These processes, which violate conservation laws normally observed 
to hold very precisely in the laboratory, are experimentally
inaccessible because they are expected to occur only at 
temperature\cite{KUZ}
or fermion densities\cite{MATV} inaccessible at 
laboratory scales.  Nor are these
processes thought to be available for study in any contemporary astrophysical
situations. 
Within the standard Big Bang model, the electroweak phase transition from the
state where electroweak symmetry is exact and the weak bosons are massless
to the spontaneously broken ground state that we see today occurred
early, when the universe was extremely hot and dense.  Here, we address
whether such extremes of temperature and/or densities can lead to this
phase transition in contemporary processes.

In the next section we examine the 
relationship between the
electroweak phase transition\cite{MATV}
in highly compressed matter and the formation
of black holes during gravitational collapse. We shall show that the
densities and temperatures achieved in gravitational collapse of stellar
size masses do not approach those needed for the electroweak phase
transition. 

In view of the potential importance of finding an experimental system in
which the electroweak phase transition might be observed today, we examine a
speculative possibility. Sonoluminescence\cite{CRUM} in water is an acoustical process
which focuses mechanical energy from macroscopic dimensions down to nearly
atomic dimensions. (The most likely mechanism is cavitation followed by
forced collapse of the bubble. The speed of the collapsing bubble wall
exceeds the speed of sound in the water vapor within the bubble and a shock
wave is formed, which focuses a great deal of energy at a point.  In a 
subsequent section of this paper we suggest an alternative scenario applicable
to sonoluminescence in neutron stars.) If
sonoluminescence should somehow occur in a neutron star as a result, say, of
a star quake or supernova core bounce, we could have focusing of energy over
a stellar volume, down, perhaps, to nearly nuclear dimensions; a volume
scale factor of the order of 10$^{30}$over the laboratory scale 
sonoluminescence, (as well as
a large scale up in energy). We calculate the minimum energy needed to
overcome lepton cooling and to compress matter to the electroweak scale. We
find that energy required is small compared to that available in a star
quake. We shall also note that sonoluminescence does not seem to have the
potential to form black holes in neutron stars.

The electroweak phase transition can be reached by way of high 
temperature or high fermion density (which is to say, high
 Fermi level) but the characteristic energy of the transition 
and the rate of baryon and lepton violation can only be 
determined by means of detailed modeling.  Several authors\cite{SHAP}
report that the characteristic energy, either temperature or
 Fermi level, is in the range 100 to 200 GeV.  
Based upon those results, we shall assume
that baryon number and lepton number violation is important above 100 GeV.
 
\section{Electroweak Transition in Gravitational Collapse}

Gravitational collapse often proceeds through a low temperature compression
of material bodies of stellar mass scale, as when mass is accreted on a
neutron star. In this case, the conditions for the electroweak transition
will not be reached for several reasons. 

If we set the Fermi level of a degenerate fermion core
(T = 0) equal to E$_c$, 
we obtain a fermion density which is typical of that
at the phase transition. If E$_c$ = 100 GeV,
($\frac nV)_c=4\times10^6$ /fm$^3$,
$\rho _c=5 \times 10^{23}$ gm/cm$^3$. If we draw this density as a line on a
curve which portrays density vs radius for astronomical objects\cite{SHAPI}, 
we find
that the electroweak density line intersects the black hole line where
the radius of the object equals its Schwarzschild radius, when
\begin{equation}
\frac{E_c^4}{(3\pi ^2)}=\frac 1{2\pi GR^2}  \label{Critical}
\end{equation}

This corresponds to a mass of M$_c\sim 3\times 10^{26}$kg 
$\sim $ 10$^{-4}M_{sun} $ and a radius of R$_c\sim $ 50~cm. 
An object with such a
low mass has few natural processes via which to achieve such high densities.

Moreover, the critical density cannot be reached in isothermal compression
since compression leads to copious $W$ boson production at a threshold 
equal to the $W$ mass, which prevents the Fermi level from reaching electroweak
scale because of lepton cooling; energy expended in further compression 
is radiated away into the body of the neutron star by the leptons 
deriving from the boson decay.

\section{Sonoluminescence in Neutron Stars}

Since sonoluminescence in water only partly understood\cite{CRUM} 
at this time, we
cannot build a convincing model of the same phenomenon in a neutron star.
However, while the occurrence of a phenomenon analogous to sonoluminescence
in a neutron star is strictly speculative, such a phenomenon may be 
{\it the only (contemporary) process} which yields any 
chance of achieving electroweak phase
transitions.  In the next section, we propose a qualitative
scenario to argue that it may be possible to achieve
sonoluminescence in a neutron star under some circumstances.

In neutron stars, interparticle energies are negligible compared to the
thermodynamic energy of the Fermi gas as well as compared to the gravitational
energy. Waves can therefore propagate without loss until the energy is very
high. Possible driving forces might be infalling matter, star
quakes\cite{SHAPI} (which
have equatorial symmetry and release on the order of 10$^{42}$ ergs), and,
most importantly, residual energy from supernova explosions\cite{MAY} (the
approximately spherically symmetric 'bounce' energy following the collapse
with an release of the order $10^{52}$ ergs).
We conjecture that the focusing of this energy into a very small volume
at the center of a neutron star may result in compression heating of the
material. 

We have already noted that isothermal, T = 0, compression cannot
achieve the critical density.
We shall next argue that adiabatic compression can, in principle
achieve temperatures as high as $E_c$. If the temperature of
the fermions exceeds 100 GeV, $W$ bosons are formed
producing prompt leptons. 
The absorption length of leptons of energy $\sim 100 \ GeV$
is approximately 
\[
\Lambda _W\simeq \frac 1{(\frac nV)_W \ \sigma (100 \ GeV)}\simeq
10^{-9} {\it m}. 
\]
where ($\frac nV)_W\simeq 5.10^6/fm^3$ is the density of relativistic 
fermions at temperature $\stackrel{\sim}{>}$ 100 GeV.
Assuming a spherical volume being compressed, the
radius must be very large compared to $\Lambda _W$ in order that the lepton
cooling be suppressed. That is

\begin{equation}
R_W\gg \Lambda _W.  \label{Radius}
\end{equation}

The condition \ref{Radius} implies that a minimum acoustical energy input, $E$,
is needed to achieve adiabatic compression beyond the W threshold; $E\gg
\;E_W,$ where $E_W$ is the order of magnitude energy of a sphere of quarks
at the $W$ threshold (with $N_W$= the number of fermions in the sphere ).

\begin{eqnarray*}
E_W &\simeq &\frac{4\pi }3(\frac nV)\Lambda _W^3  (100 \ GeV) 
\sim 10^{24} {\it erg}. 
%E_W &\simeq &N_WM_W\sim 10^{23}{\it erg}
\end{eqnarray*}

Beyond the W phase transition there are several species of particles present.
We assume thermodynamic equilibrium and equal populations of all quarks and
leptons produced below 100 GeV, implying that $N_s\sim 28$ (We count all
leptons and quarks with lifetimes long compared to 10$^{-10}$ seconds.) The
density of each species will be the same as given above. The absorption
length 
\[
\Lambda _c=\frac 1{N_s(\frac nV)_c\sigma (\frac 12E_c)} 
\]
will be smaller than $\Lambda _W,$ both because the cross section is rising
with energy and because the number density is higher. Therefore, lepton
cooling decreases with increasing compression beyond the W threshold.
We thus conclude that via adiabatic compression, temperatures as
high as $\sim 100$~GeV may possibly be achieved; the minimum energy
required should substantially exceed $E_{\min}$ given by, 
\begin{equation}
E_{\min }\simeq 10^{23}-10^{25} {\it erg},  \label{Energy}
\end{equation}
which is very small compared to the star quake\cite{SHAP} and 
supernova\cite{MAY} energies
mentioned above.
We conclude from these dimensional arguments that the adiabatic conditions
can be met.

Since the ratio of energy released by changing the initial population of
neutrons into leptons to
the electroweak ignition energy is small, $m_n/E_c\simeq 1/100$, sustained
burning is not possible. Therefore, once the threshold for the electroweak
phase transition has been exceeded, the sphere contains mainly high energy
leptons which escape into the body of the star in a time of the order of 
$R_W/c$. The neutrinos diffuse out of the star when they have degraded
in energy such that $\Lambda \sim R_{star}=10km.$  Setting
\[
\Lambda \approx 10km = \frac{1}{\left (\frac{n}{V} \right )_{ns} \sigma}
\]
where ($\frac nV)_{ns}\simeq 6 \times 10^{37}/cm^3$ is the number density in a
typical neutron star and $\sigma \sim 9 \times 10^{-42} \ cm^2 \left (
\frac{E}{10 \ MeV} \right )^2$ is the low energy 
neutrino - nucleon cross section, E$_{esc}$ is found to be of the order of 1
MeV.
We note that the 
energy radiated from the star as neutrinos will be large compared to 
$E_{\min }$ in eq.(\ref{Energy}) but, of course, smaller than the driving
energy.\footnote{
Poplitz\cite{POP} has shown that magnetic fields of the order of 10$^{33}$Gauss are
required for the electroweak transition. Therefore, typical neutron star
magnetic fields will not affect our results.}

Finally, we ask whether the compression due to sonoluminescence could
result in the formation of a blackhole.
The order of magnitude of the energy required to achieve the electroweak
threshold and the black hole threshold is $\sim M_c\cdot A_0 \cdot (100GeV)\sim
10^{52}erg$ ($A_0$ is Avogadro's number). Consistent with the lepton cooling
arguments, above, we could scale the radius down by no more than about 
$10^{-5}$, but to achieve the black hole condition the density would scale up
by $10^{10}$, and black hole formation would still require of the order 
of $10^{47}erg$. Therefore, we conclude that black hole production by
sonoluminescence is unlikely.

\section{A Scenario for Sonoluminescence }

In the previous sections we treated sonoluminescence in a neutron star as 
a speculation.  In this section, we introduce a scenario which serves to make
our speculation somewhat plausible.  
We suppose that the energy of the bounce from
neutronizaton in a supernova explosion is deposited in an acoustic field in
the neutron star (or proto-neutron star) and is propagating centrally inward 
carrying a kinetic energy of the order of\cite{MAY}
\[
E_{Bounce} = 10^{52} \ ergs.
\]
When the energy density of the acoustic field at some radius,
$R \ll R_{star}$, exceeds the threshold for the production of a new species of
particle, there is a brief period
of time during which the energy is used to create new species of particles;
during this interval, the medium becomes ``soft'', and the disturbance
cannot propogate as rapidly.
After this brief interval, the local Fermi levels of all fermion
final states fill up to the threshold energy, the particle production
process saturates, the speed of sound (at R) resumes its nominal value,
$c_s=c/\sqrt{3}$,
%which is slightly less than the speed of light, 
and the active region of 
particle production moves inward to a smaller radius.  Thus,
a spherical discontinuity forms and propagates inward, driven by the
acoustic field impinging from the body of the star.  The 
front edge of this discontinuity 
travels at a velocity
lower than that of sound and can, therefore accumulate energy from
the acoustic wave.  
The acoustic energy absorbed in a thin shell at R goes partly into production 
of particles and we propose, partly into increased energy of 
the discontinuity. Assuming that the acoustic wave is not {\it all} reflected
back at the discontinuity, it is plausible that 
the inward focusing discontinuity 
continues to increase in energy density while encountering a series of 
higher and higher thresholds for 
particle production at smaller and smaller radii.  The density increases due 
to compression of the initial material present in the star, plus the
compression of newly formed Fermi seas.

The issue then is whether the acoustic wave can drive the discontinuity
to the centre of the star, where by adiabatic compression, a temperature
above 100~GeV, and the accompanying effects discussed previously might result.
Since an acoustic wave would take $\sim 10^{-5}s$ to traverse a star of
radius $\sim 1 km$, we believe that it is reasonable to suppose
that the discontinuity gets driven to the center provided that the initial
disturbance (perhaps the energy of the bounce from neutronization) that causes
the energy to be deposited into the acoustic field lasts substantially
longer. Here we remark that the typical time scale for the initial neutrinos
produced dynamically in supernova collapse is $\sim 10 ms \gg 10^{-5}s$
mentioned above. Finally, we remark that the energy $E_{Bounce}$ is
truly enormous, so that even if the bulk of it is reflected at the
discontinuity, the mechanism that we propose would be energetically
possible. Of course, whether or not it will actually occur will depend
on many detailed issues beyond the scope of the present analysis. 

\section{Concluding Remarks}
 
We have noted that on a black hole phase diagram, the line portraying the
density of matter at the electro-weak transition intersects the black 
hole curve at a point, eq(\ref{Critical}).  
The right hand side 
this equation depends only upon gravity 
while the left hand side depends only upon
electroweak physics, so that this provides an intriguing
connection between these areas of
physics which have no common fundamental 
theoretical basis at the present time.
%The mass and radius values are points of 
%contact between two parts of physics 
     
We have shown that in the normal course of gravitational collapse, the 
electroweak density will not be reached.  However, we have suggested 
a mechanism, sonoluminescence in a neutron star, 
which might achieve the electroweak phase transition. While this
mechanism is only a speculation on our part, 
it is of some interest since this may well be the only contemporary process  
where the electroweak phase transition would be of relevance.
We have also 
presented a very simple hydrodynamic scenario to support the 
plausibility of our speculation of sonoluminescence.
Should the sonoluminescence process occur during star quake or supernova, 
(with or without attainment of the electroweak threshold), the result 
is a pulse of low energy neutrinos  (E $\sim$ MeV) 
of all flavors which diffuse from the star.
The neutrino flux from such a process at 
10 kpc would not register any events with even the low-threshold solar 
neutrino detectors such as Borexino\cite{ARP} under construction at Grand
Sasso and Hellaz\cite{LAU} in the design stage at present.
If such an event occurs as close as 1 kiloparsec, 
then it may be possible to see 10 to 20 neutrino events of a time scale 
$10^{-3}$ sec due to 
sonoluminescence in a detector such as Borexino.
The events would stand out above backgrounds (due to solar neutrinos) and be
distinguished from the neutronization burst events by being lower in energy
(average energy being close to 1 MeV rather than 10 MeV for neutronization
$\nu's$) and being later in time.
The secondary thermal neutrinos are expected to be emitted as usual; 
these were
observed from SN1987A but due to the high thresholds of the detectors no
conclusion can be drawn about possible low energy neutrinos from sonolominescence.  Future low threshold ($E_{th} <$ 1 MeV) high volume detectors would be
able to study this question in detail.
    We have shown that the electro-weak phase transition can only occur if the
 energy of the sonoluminescence discontinuity is large compared to 
$10^{25}ergs$, which is necessary to overcome cooling by prompt 
leptons when the energy per particle rises above the W boson 
production threshold.  The additional energy released
in the phase transition is expected to be less than about 1 GeV for each 
100 GeV of
sonoluminescence input energy.  Expansion of the electroweak phase to 
consume a large part of the star is impossible; a spherical detonation 
wave cools faster by adiabatic
expansion than it can heat up by burning 
nucleons at neutron star densities.
    
In summary, sonoluminescence in a neutron star appears possible and has 
interesting consequences, but the details depend upon many factors beyond
the scope of our analysis.

\section{Acknowledgement}

We wish to thank V. Rubakov for useful discussions and for a careful 
reading of the manuscript.

\end{document}